%% file: DARS-SOCIAL.tex
\documentclass[10pt,conference]{IEEEtran}
\IEEEoverridecommandlockouts
\usepackage[hidelinks]{hyperref}
\usepackage{booktabs} 
\usepackage{amsmath,amsfonts,amssymb,epsfig,epstopdf,url,array}

\usepackage{amsthm}
\usepackage{color,soul}
\usepackage{algorithmic}
\usepackage{balance}
\usepackage[ruled]{algorithm2e}
\usepackage{stmaryrd}
\usepackage{centernot}
\usepackage{tikz}
\usepackage{booktabs}
\usepackage{pdfpages}
\usepackage{graphicx}
\usepackage{subfigure}
\usepackage[font=small,skip=8pt]{caption}
\usepackage{tabularx}
\usepackage{comment}
\usepackage{multirow}
\usepackage{multicol}  
\usepackage{bigstrut}
\usepackage{pdflscape}
\usepackage{rotating}
\usepackage{float}
\usepackage{lipsum}

\usepackage{enumitem}
\usetikzlibrary{plotmarks,shapes,arrows,chains,hobby,backgrounds,calc,trees}
\usepackage{makecell}
\usepackage{dblfloatfix}    
\usepackage{pdfcomment} 
\usepackage{etoolbox}
\usepackage{colortbl}
\usepackage{balance}

\definecolor{lightgray}{gray}{0.9}
\theoremstyle{definition}

\theoremstyle{definition}
\newtheorem{mydef}{Definition}
\theoremstyle{definition}

\AtBeginEnvironment{align}{%
	\fontsize{10}{11}%
}

\begin{document}

\title{Human Values in Software Release Planning}

\author{\IEEEauthorblockN{Davoud Mougouei, Aditya Ghose, Hoa Dam}
\IEEEauthorblockA{\textit{School of Computing and Information Technology} \\
\textit{University of Wollongong}\\
Sydney, Australia \\
\{Davoud, Aditya, Hoa\}@uow.edu.au}
\and
\IEEEauthorblockN{David Powers}
\IEEEauthorblockA{\textit{College of Science and Engineering}} 
\textit{Flinders University}\\
Adelaide, Asutralia \\
David.Powers@flinders.edu.au}

\maketitle
\thispagestyle{plain}
\pagestyle{plain}

\begin{abstract}
\input{abstract}
\end{abstract}


\input{introduction}
\input{modeling}

\input{selection}
\input{summary}

\balance
\bibliographystyle{IEEEtran}
\bibliography{IEEEabrv,ref}

\end{document}

%% file: abstract.tex
Software products have become an integral part of human lives, and therefore need to account for human values such as privacy, fairness, and equality. Ignoring human values in software development leads to biases and violations of human values: racial biases in recidivism assessment and facial recognition software are well-known examples of such issues. One of the most critical steps in software development is Software Release Planning (SRP), where decisions are made about the presence or absence of the requirements (features) in the software. Such decisions are primarily guided by the economic value of the requirements, ignoring their impacts on a broader range of human values. That may result in ignoring (selecting) requirements that positively (negatively) impact human values, increasing the risk of value breaches in the software. To address this, we have proposed an \textit{Integer Programming} approach to considering human values in software release planning. In this regard, an \textit{Integer Linear Programming} (ILP) model has been proposed, that explicitly accounts for human values in finding an ``optimal'' subset of the requirements. The ILP model exploits the algebraic structure of fuzzy graphs to capture dependencies and conflicts among the values of the requirements. 

%% file: introduction.tex
\section{Introduction}
\label{ch_so_introduction}

Today's software products collect and analyze our biological data, personalize recommendations based on our usage patterns and emotions, and monitor and influence our behavior~\cite{osuna2021toward}. That gives software products a great deal of power and influence over humans. But ``with great power comes great responsibility.'' Studies have shown that considering human values such as privacy, fairness, and equality, in developing software, is essential for mitigating the potential risks it can pose to the society~\cite{vinuesa2020role}. Ignoring human values, on the other hand, may lead to privacy violations, algorithmic biases, and widening socio-economic inequalities. One of the most important stages in software development is software release planning, where decisions are made as to what requirements (features) should be included in the software. Ignoring human values at such a critical point may lead to breaching those values in the software, leading to user dissatisfaction~\cite{mougouei2018operationalizing}, financial loss, and reputation damage~\cite{perera2020study,mougouei2020engineering}. 

Several examples of violating human values in software have been reported in recent years~\cite{khademi2020algorithmic,bellamy2019think}: racial biases in recidivism assessment, facial recognition software, and photo tagging~\cite{bellamy2019think} are of the most commonly known examples~\cite{khademi2020algorithmic}. Among these, perhaps one of the most tragic cases is the suicide of the British teenager Molly Russell~\cite{molly.2019}, attributed to Instagram's personalization algorithms, which flooded Molly's feed with self-harm images without considering the impact of such personalization on her health. In response, Instagram added features that prevent personalizations on self-harm. Had such features and their impacts on health been evaluated more carefully in the earlier stages of software development, Molly's suicide could have possibly been prevented. Hence, it is important to account for human values as early as in software release planning, where decisions are made about the requirements (features) of the software, based on their values. 

\begin{figure*}[!htbp]
	\centering\includegraphics[scale=0.8]{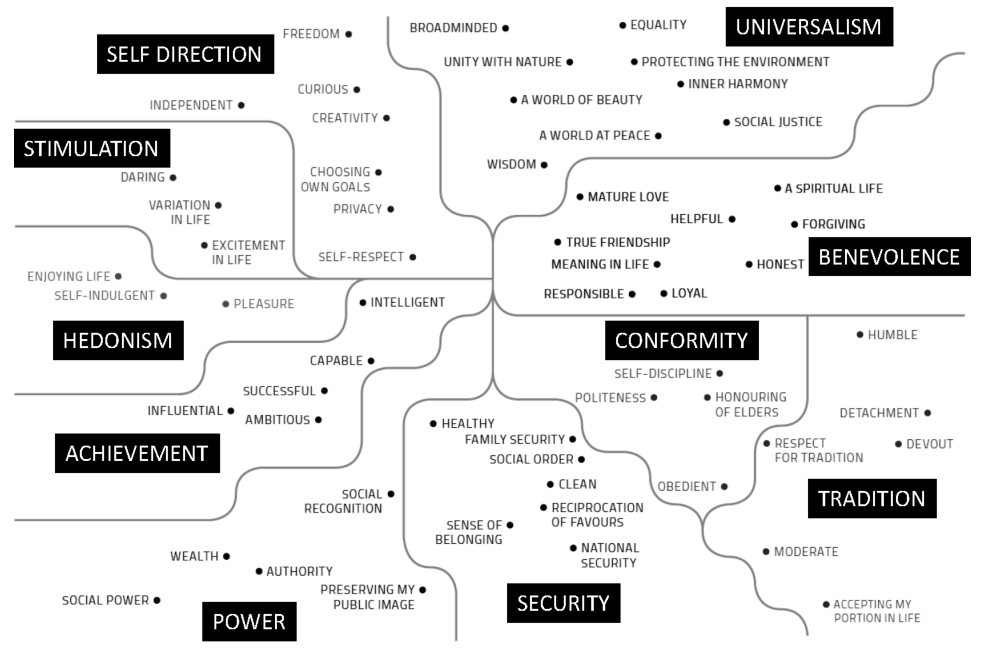}
	\caption{%
		A map of human values based on Schwartz theory of basic values~\cite{schwartz1987toward}.
	}%
	\label{fig_map}
\end{figure*}

%
Regardless of the approach used for release planning, the process involves selecting an ``optimal'' subset of the requirements. Release planning thus is modeled as different variations of Binary Knapsack Problem (BKP), for which several mathematical formulations have been devised~\cite{MOUGOUEI2020113748,mougouei2017integer}. These formulations can be solved, exactly, using advanced solvers such as IBM CPLEX~\cite{mougouei2020dependency} or be approximated using meta-heuristics~\cite{zhang2018empirical}. Nonetheless, the existing software release planning models have primarily focused on maximizing the economic value of software, ignoring a broader range of human values (Figure~\ref{fig_map})~\cite{mougouei2018operationalizing}, that are also important and need to be taken into account. Moreover, the extant release planning models do not take into account the fact that the value of a software requirement may depend on the presence or absence of other requirements in the optimal set. Ignoring such \textit{Value Dependencies} increases the risk of value loss through ignoring (selecting) requirements that positively (negatively) influence the values of other requirements~\cite{sangwan2020optimization,MOUGOUEI2020113748,mougouei2019dependency,mougouei2016factoring,mougouei2017dependency,mougouei2017modeling}. It is imperative that value dependencies are not limited to the economic value; they extend to a broader range of human values~\cite{shi2017new,mougouei2018operationalizing}. As observed by Carlshamre \textit{et al.}~\cite{carlshamre_industrial_2001}, requirement dependencies in general and value dependencies in particular are \emph{fuzzy}~\cite{carlshamre_industrial_2001} in the sense that the strengths of those dependencies are imprecise and vary~\cite{dahlstedt2005requirements,ngo_wicked_2008,ngo2005measuring,carlshamre_industrial_2001} from large to insignificant~\cite{wang_simulation_2012} in real-world projects. Hence, it is important to consider not only the existence but the strengths of value dependencies and the imprecision of those dependencies in software projects.

In this paper, we propose an integer programming approach to considering human values in software release planning. The proposed approach comprises an \textit{Integer Linear Programming} (ILP) model, that finds an optimal subset of the software requirements with the highest economic worth subject to a set of constraints concerning human values, i.e., \textit{Value Constraints}. The value constraints of a software project ensure that the user values are not violated in the selected subset of the requirements. We have further exploited the algebraic structure of fuzzy graphs to capture dependencies and conflicts among the values of the requirements. In this regard, we have proposed an extension of fuzzy graphs named \textit{Typed Value Graphs} (TVGs), that allow for capturing the strengths and qualities (positive/negative) of value dependencies. 

%% file: modeling.tex
\section{Modeling Dependencies among Values}
\label{ch_soc_modeling}
\hypertarget{ch_soc_modeling}{ }	


 
This section presents typed value graphs (TVGs) for modeling value dependencies among the requirement. A TVG of type $t$ is defined by Definition~\ref{def_ch_soc_vdg}, where $t$ denotes a specific human value, e.g., privacy, fairness, and economic worth. Multiple TVGs can be created for a software project, each capturing the value dependencies of a specific type. 

\begin{mydef}
	\label{def_ch_soc_vdg}
	\textit{Typed Value Graph}. A type $t$ TVG is a signed directed fuzzy graph $G_t=(R,\sigma_t,\rho_t)$ , where the requirement set $R:\{r_1,...,r_n\}$ form the graph nodes. Also, the qualitative function $\sigma_t \rightarrow \{+,-,\pm\}$ and the membership function $\rho_t: (r_i,r_j)\rightarrow [0,1]$ specify the quality and the strength of an explicit type $t$ value dependency from $r_i$ to $r_j$ respectively. $\rho_t(r_i,r_j)=0$, $\sigma_t(r_i,r_j)=\pm$ specifies the absence of an explicit type $t$ value dependency from $r_i$ to $r_j$. 
\end{mydef}


%
Definition~\ref{def_ch_soc_vdg_valuedepndencies} provides a comprehensive definition of value dependencies that includes both explicit (direct) and implicit (indirect) value dependencies of different types. 

\begin{mydef}
	\label{def_ch_soc_vdg_valuedepndencies}
	\textit{Typed Value Dependencies}. 
	A type $t$ value dependency in a TVG $G_t=(R,\sigma_t,\rho_t)$ is defined as a sequence of the requirements $d_i:\big(r(0),...,r(k)\big)$ such that $\forall r(j) \in d_i$, $1 \leq j \leq k$ we have $\rho_t\big(r(j-1),r(j)\big) \neq 0$. $j\geq 0$ is the sequence of the $j^{th}$ requirement (node) denoted as $r(j)$ on the dependency path. A consecutive pair $\big(r(j-1),r(j)\big)$ specifies an explicit value dependency. 
\end{mydef}

\vspace{-0.5cm}
\begin{align}
\label{Eq_ch_soc_vdg_strength}
&\forall d_i:\big(r(0),...,r(k)\big): \rho_t(d_i) = \bigwedge_{j=1}^{k}\text{ }\rho_t\big(r(j-1),r(j)\big) \\
\label{Eq_ch_soc_vdg_quality}
&\forall d_i:\big(r(0),...,r(k)\big): \sigma_t(d_i) = \prod_{j=1}^{k}\text{ }\sigma_t\big(r(j-1),r(j)\big)
\end{align}

Equation (\ref{Eq_ch_soc_vdg_strength}) computes the strength of type $t$ value dependency $d_i:\big(r(0),...,r(k)\big)$ by finding the strength of the weakest of the $k$ explicit type $t$ dependencies on $d_i$. Fuzzy operator $\wedge$ denotes Zadeh's AND (minimum) operation~\cite{zadeh_fuzzysets_1965,mougouei2013fuzzy}. Also, the quality (positive or negative) of a type $t$ value dependency $d_i:\big(r(0),...,r(k)\big)$ is calculated by the qualitative serial inference~\cite{de1984qualitative,wellman1990formulation,kusiak_1995_dependency} of (\ref{Eq_ch_soc_vdg_quality}) as shown in Table~\ref{table_ch_soc_inference}. 
\begin{table}[!htb]
	\caption{Qualitative serial inference in a type $t$ TVG.}
	\label{table_ch_soc_inference}
	\centering
	\input{table_inference}

\end{table}

Let $D_t=\{d_1,d_2,..., d_m\}$ be the set of all type $t$ value dependencies from $r_i \in R$ to $r_j \in R$ in a type $t$ TVG $G_t=(R,\sigma_t,\rho_t)$, where the positive and negative dependencies can simultaneously exist from $r_i$ to $r_j$. The strength of all positive value dependencies of type $t$ from $r_i$ to $r_j$ is denoted by $\rho_{t}^{+\infty}$ and calculated by (\ref{Eq_ch_soc_ultimate_strength_positive}), that is to find the strength of the strongest positive dependency~\cite{rosenfeld_fuzzygraph_1975} from $r_i$ to $r_j$. Fuzzy operators $\wedge$ and $\vee$ denote Zadeh's fuzzy AND (minimum) and fuzzy OR (maximum) operators~\cite{zadeh_fuzzysets_1965} respectively. Similarly, the strengths of all negative value dependencies of type $t$ from $r_i$ to $r_j$ is denoted by $\rho_{t}^{-\infty}$ and computed by (\ref{Eq_ch_soc_ultimate_strength_negative}).


\begin{align}
\label{Eq_ch_soc_ultimate_strength_positive}
&\rho_t^{+\infty}(r_i,r_j) = \bigvee_{d_m\in D_t, \sigma_t(d_m)=+} \text{ } \rho_t(d_m) \\[1pt]
\label{Eq_ch_soc_ultimate_strength_negative}
&\rho_t^{-\infty}(r_i,r_j) = \bigvee_{d_m\in D_t, \sigma_t(d_i)=-} \text{ } \rho_t(d_m) \\[1pt]
\label{Eq_ch_soc_influence}
&I_{i,j,t} = \rho_t^{+\infty}(r_i,r_j)-\rho_t^{-\infty}(r_i,r_j) 
\end{align}


In a type $t$ TVG, $\rho_t^{+\infty}(r_i,r_j)$ or $\rho_t^{-\infty}(r_i,r_j)$ can be computed in polynomial time, $O(n^3)$, using Floyd-Warshall algorithm. We have demonstrated this for a simple value graph, with only one type (economic value), in our earlier work~\cite{MOUGOUEI2020113748}. It is clear that repeating the computations for different values types does not change the order of complexity. The overall strengths of all type $t$ value dependencies from $r_i$ to $r_j$ is referred to as the \textit{Influence} of $r_j$ on the type $t$ value of $r_i$ and denoted by $I_{i,j,t}\in[-1,1]$. $I_{i,j,t}$, as given by (\ref{Eq_ch_soc_influence}), is calculated by subtracting $\rho_t^{-\infty}(r_i,r_j)$ from $\rho_t^{+\infty}(r_i,r_j)$. A positive (negative) values for $I_{i,j,t}$  implies that $r_j$ positively (negatively) influences the type $t$ value of $r_i$.

%% file: table_inference.tex
\resizebox {0.35\textwidth }{!}{
	\begin{tabular}{cc|ccc}
		\toprule[1.5pt]
		\multicolumn{2}{r|}{\multirow{2}[1]{*}{ $\sigma_t\big(r(j-1),r(j),r(j+1)\big)$}} &
		\multicolumn{3}{c}{$\sigma_t\big(r(j),r(j+1)\big)$}
		\\
		\multicolumn{2}{r|}{} &
		$+$ &
		$-$ &
		$\pm$
		\bigstrut[b]\\
		\hline
		\multicolumn{1}{c}{\multirow{3}[1]{*}{$\sigma_t\big(r(j-1),r(j)\big)$}} &
		$+$ &
		$+$ &
		$-$ &
		$\pm$
		\bigstrut[t]\\
		\multicolumn{1}{c}{} &
		$-$ &
		$-$ &
		$+$ &
		$\pm$
		\\
		\multicolumn{1}{c}{} &
		$\pm$ &
		$\pm$ &
		$\pm$ &
		$\pm$
		\\
    \bottomrule[1.5pt]
	\end{tabular}%
	}

%% file: selection.tex
\section{The Integer Linear Programming Model}
\label{ch_soc_selection}
\hypertarget{ch_soc_selection}{ }	



 
 Equations (\ref{eq_soc_ilp})-(\ref{eq_soc_ilp_c11}) give an integer linear programming (ILP) model that finds an optimal subset of the requirements with the highest economic worth subject to a set of constraints concerning human values, i.e., \textit{Value Constraints}. The model uses the algebraic structure of typed value graphs (TVGs) introduced in Section~\ref{ch_soc_modeling} for capturing value dependencies among the requirements. In (\ref{eq_soc_ilp})-(\ref{eq_soc_ilp_c11}), $x_i$ is a selection variable denoting whether a requirement $r_i$ is selected ($x_i=1$) or ignored ($x_i=0$). Also, $\theta_{i,t}$ in (\ref{eq_soc_ilp_c4}) specifies the penalty of a type $t$ value of requirement $r_i$, that is the extent to which the type $t$ value of $r_i$ is impacted by ignoring (selecting) requirements with positive (negative) influences on the type $t$ value of $r_i$. $T$ in (\ref{eq_soc_ilp})-(\ref{eq_soc_ilp_c11}) specifies the total number of the value types including the economic value ($T=58$ as in Figure~\ref{fig_map}). Moreover, $\theta_{i,t}$ in (\ref{eq_soc_ilp_c4}) gives the simplified algebraic expression of the penalty in (\ref{eq_ch_soc_penalty}). $I_{i,j,t}$, as given by (\ref{Eq_ch_soc_influence}), is the influence of $r_j$ on the type $t$ value of $r_i$. 

\begin{align}
\label{eq_ch_soc_penalty}
\nonumber
\theta_{i,t}= &\displaystyle \bigvee_{j=1}^{n} \bigg(\frac{x_j\big(\lvert I_{i,j,t} \rvert-I_{i,j,t}\big) + (1-x_j)\big(\lvert I_{i,j,t}\rvert+I_{i,j,t}\big)}{2}\bigg)= \\ \nonumber
&\displaystyle \bigvee_{j=1}^{n} \bigg(\frac{\lvert I_{i,j,t} \rvert + (1-2x_j)I_{i,j,t}}{2}\bigg), \\  & i\neq j = 1,...,n,\text{ }t=1,...,T 
\end{align}

As stated earlier, $v_{i,1}$ in (\ref{eq_soc_ilp})-(\ref{eq_soc_ilp_c11}) denotes the economic value of a requirement $r_i$ and $E(v_{i.t})$, $t \in \{2,...,n\}$ denotes the expected type $t$ value of a requirement $r_i$. Similarly, in all other variables/parameters in (\ref{eq_soc_ilp})-(\ref{eq_soc_ilp_c11}), $t=1$ denotes the variables/parameters related to the economic value (Wealth) while $t=\{2,...,T\}$ specify the variables/parameters related to other values in Figure~\ref{fig_map}. The expected values of the requirements are used in the proposed ILP model to account for the fact that a particular requirement (feature) may not be perceived equality valuable by different users.

\begin{align}
\label{eq_soc_ilp}
&\text{Maximize }  \sum_{i=1}^{n} x_i E(v_{i,1}) - y_{i,1} E(v_{i,1}) \\[2pt]
\label{eq_soc_ilp_c1}
&\text{Subject to} \sum_{i=1}^{n} c_i x_i \leq b\\[2pt]
\label{eq_soc_ilp_c2}
& \begin{cases}
x_i \le x_j  & r_j \text{precedes } r_i \\[2pt]
x_i \le 1-x_j& r_i \text{ conflicts with } r_j,\text{ }i\neq j= 1,...,n
\end{cases}\\[5pt]
\label{eq_soc_ilp_c3}
& \sum_{i=1}^{n} x_i E(v_{i,t}) - y_{i,t} E(v_{i,t}) \geq \beta_t,\hspace{1.2cm} t=2,...,T \\[2pt]
\label{eq_soc_ilp_c4}
& \theta_{i,t}\geq \bigg(\frac{\lvert I_{i,j,t} \rvert + (1-2x_j) I_{i,j,t}}{2}\bigg), \\ \nonumber
& \hspace{3.51cm}  i\neq j = 1,...,n,\text{ }t=1,...,T \\[2pt]
\label{eq_soc_ilp_c5}
& -g_i \leq x_i \leq  g_i,\hspace{3.6cm} i=1,...,n\\[2pt]
\label{eq_soc_ilp_c6}
& 1-(1-g_i) \leq x_i \leq 1+(1-g_i), \hspace{1.05cm}i=1,...,n\\[2pt]
\label{eq_soc_ilp_c7}
& -g_i \leq y_{i,t} \leq g_i, \hspace{1.55cm} i=1,...,n,\text{ }t=1,...,T\\[2pt]
\label{eq_soc_ilp_c8}
& -(1-g_i)\leq(y_{i,t}-\theta_{i,t}) \leq (1-g_i),\\ \nonumber
& \hspace{4.1cm}i=1,...,n,\text{ }t=1,...,T \\[2pt]
\label{eq_soc_ilp_c9}
&\text{ } 0 \leq y_{i,t} \leq 1, \hspace{2.05cm} i = 1,...,n,\text{ }t=1,...,T\\[2pt]
\label{eq_soc_ilp_c10}
&\text{ } 0 \leq \theta_{i,t} \leq 1,\hspace{2.06cm} i = 1,...,n,\text{ }t=1,...,T\\[2pt]
\label{eq_soc_ilp_c11}
& \text{ }x_i,g_i \in \{0,1\},\hspace{3.65cm} i=1,...,n
\end{align}

The objective function (\ref{eq_soc_ilp}) aims to optimize the economic value of the selected requirements subject to (\ref{eq_soc_ilp_c1})-(\ref{eq_soc_ilp_c11}). Constraint (\ref{eq_soc_ilp_c1}) ensures that the total cost of the requirements will not exceed the budget $b$. Also, (\ref{eq_soc_ilp_c2}) accounts for the precedence dependencies among the requirements and the value implications of those dependencies, which may impact the value types. Precedence dependencies mainly include the requirement dependencies of type \textit{Requires} (\textit{Conflicts-with}), where one requirement intrinsically requires (conflicts with) another. The set of the Value Constraints (\ref{eq_soc_ilp_c3}) ensures that the minimum amounts (lower-bounds) of the expected type $t$ values are satisfied for the selected requirements. $\beta_{t}$ in (\ref{eq_soc_ilp_c3}) denotes the required lower-bound for the expected type $t$ value of the requirements. $\beta_t$ can be adjusted to reflect the preferences of the stakeholders for different human values, and reconcile potential conflicts among the values when satisfying one value type conflicts with another. For a given requirement $r_i$, in (\ref{eq_soc_ilp})-(\ref{eq_soc_ilp_c11}) we have either $a:(x_i=0,y_{i,t}=0), t= 1,...,T$, or $b:(x_i=1,y_{i,t}=\theta_{i,t}), t=1,...,T$ occur. To capture the relation between $\theta_{i,t}$ and $y_{i,t}$ in a linear form, we have made use of an auxiliary variable $g_i=\{0,1\}$ and (\ref{eq_soc_ilp_c5})-(\ref{eq_soc_ilp_c11}). As such, we have either $(g_i=0) \rightarrow a$, or $(g_i=1) \rightarrow b$. The selection model (\ref{eq_soc_ilp})-(\ref{eq_soc_ilp_c11}) therefore, is a linear model as it has a linear objective function with linear inequality constraints. 

%% file: summary.tex
\section{Summary}

In this paper, we proposed an integer programming approach to considering human values in software release planning. At the heart of the proposed approach is an integer linear programming (ILP) model that takes into account not only the economic worth of the software requirements but also their impacts on other human values such as fairness, privacy, and equality. The proposed model finds an ``optimal'' subset of software requirements with the highest economic value subject to satisfying a set of constraints concerning human values (Value Constraints). The value constraints ensure that the selected subset of the requirements does not violate the values of the users. We have further proposed typed value graphs (TVGs), a variation of fuzzy graphs, to capture the dependencies among the values of the requirements. The work
can be extended in several directions. That includes using the proposed approach for release planning of a real-world project as well as measuring the effectiveness of the proposed approach in considering human values through user surveys.

%% file: DARS-SOCIAL.bbl
\begin{thebibliography}{10}
\providecommand{\url}[1]{#1}
\csname url@samestyle\endcsname
\providecommand{\newblock}{\relax}
\providecommand{\bibinfo}[2]{#2}
\providecommand{\BIBentrySTDinterwordspacing}{\spaceskip=0pt\relax}
\providecommand{\BIBentryALTinterwordstretchfactor}{4}
\providecommand{\BIBentryALTinterwordspacing}{\spaceskip=\fontdimen2\font plus
\BIBentryALTinterwordstretchfactor\fontdimen3\font minus
  \fontdimen4\font\relax}
\providecommand{\BIBforeignlanguage}[2]{{%
\expandafter\ifx\csname l@#1\endcsname\relax
\typeout{** WARNING: IEEEtran.bst: No hyphenation pattern has been}%
\typeout{** loaded for the language `#1'. Using the pattern for}%
\typeout{** the default language instead.}%
\else
\language=\csname l@#1\endcsname
\fi
#2}}
\providecommand{\BIBdecl}{\relax}
\BIBdecl

\bibitem{osuna2021toward}
E.~Osuna, L.-F. Rodr{\'\i}guez, and J.~O. Gutierrez-Garcia, ``Toward
  integrating cognitive components with computational models of emotion using
  software design patterns,'' \emph{Cognitive Systems Research}, vol.~65, pp.
  138--150, 2021.

\bibitem{vinuesa2020role}
R.~Vinuesa, H.~Azizpour, I.~Leite, M.~Balaam, V.~Dignum, S.~Domisch,
  A.~Fell{\"a}nder, S.~D. Langhans, M.~Tegmark, and F.~F. Nerini, ``The role of
  artificial intelligence in achieving the sustainable development goals,''
  \emph{Nature communications}, vol.~11, no.~1, pp. 1--10, 2020.

\bibitem{mougouei2018operationalizing}
D.~Mougouei, H.~Perera, W.~Hussain, R.~Shams, and J.~Whittle,
  ``Operationalizing human values in software: a research roadmap,'' in
  \emph{Proceedings of the 2018 26th ACM Joint Meeting on European Software
  Engineering Conference and Symposium on the Foundations of Software
  Engineering}.\hskip 1em plus 0.5em minus 0.4em\relax ACM, 2018, pp. 780--784.

\bibitem{perera2020study}
H.~Perera, W.~Hussain, J.~Whittle, A.~Nurwidyantoro, D.~Mougouei, R.~A. Shams,
  and G.~Oliver, ``A study on the prevalence of human values in software
  engineering publications, 2015-2018,'' in \emph{2020 IEEE/ACM 42nd
  International Conference on Software Engineering (ICSE)}.\hskip 1em plus
  0.5em minus 0.4em\relax IEEE, 2020, pp. 409--420.

\bibitem{mougouei2020engineering}
D.~Mougouei, ``Engineering human values in software through value
  programming,'' in \emph{ICSEW'20: Proceedings of the IEEE/ACM 42nd
  International Conference on Software Engineering}, 2020, pp. 133--136.

\bibitem{khademi2020algorithmic}
A.~Khademi and V.~Honavar, ``Algorithmic bias in recidivism prediction: A
  causal perspective (student abstract),'' in \emph{Proceedings of the AAAI
  Conference on Artificial Intelligence}, vol.~34, no.~10, 2020, pp.
  13\,839--13\,840.

\bibitem{bellamy2019think}
R.~K. Bellamy, K.~Dey, M.~Hind, S.~C. Hoffman, S.~Houde, K.~Kannan, P.~Lohia,
  S.~Mehta, A.~Mojsilovic, S.~Nagar \emph{et~al.}, ``Think your artificial
  intelligence software is fair? think again,'' \emph{IEEE Software}, vol.~36,
  no.~4, pp. 76--80, 2019.

\bibitem{molly.2019}
N.~Baker. (2019)
  https://www.sbs.com.au/news/molly-russell-instagram-bans-graphic-self-harm-images-after-suicide-of-uk-teen.

\bibitem{MOUGOUEI2020113748}
D.~Mougouei and D.~M. Powers, ``Dependency-aware software requirements
  selection using fuzzy graphs and integer programming,'' \emph{Expert Systems
  with Applications}, p. 113748, 2020.

\bibitem{mougouei2017integer}
D.~Mougouei, D.~M.~W. Powers, and A.~Moeini, ``An integer linear programming
  model for binary knapsack problem with dependent item values,'' in \emph{AI
  2017: Advances in Artificial Intelligence: 30th Australasian Joint
  Conference, Melbourne, VIC, Australia, August 19--20, 2017, Proceedings},
  vol. 10400.\hskip 1em plus 0.5em minus 0.4em\relax Springer International
  Publishing, 2017, pp. 144--154.

\bibitem{mougouei2020dependency}
D.~Mougouei and D.~M. Powers, ``Dependency-aware software release planning
  through mining user preferences,'' \emph{Soft Computing}, 2020.

\bibitem{zhang2018empirical}
Y.~Zhang, M.~Harman, G.~Ochoa, G.~Ruhe, and S.~Brinkkemper, ``An empirical
  study of meta-and hyper-heuristic search for multi-objective release
  planning,'' \emph{ACM Transactions on Software Engineering and Methodology
  (TOSEM)}, vol.~27, no.~1, p.~3, 2018.

\bibitem{sangwan2020optimization}
R.~S. Sangwan, A.~Negahban, R.~L. Nord, and I.~Ozkaya, ``Optimization of
  software release planning considering architectural dependencies, cost, and
  value,'' \emph{IEEE Transactions on Software Engineering}, 2020.

\bibitem{mougouei2019dependency}
D.~Mougouei and D.~M. Powers, ``Dependency-aware release planning for software
  projects using fuzzy graphs and integer programming,'' \emph{Journal of
  Intelligent \& Fuzzy Systems}, pp. 1--15, 2019.

\bibitem{mougouei2016factoring}
D.~Mougouei, ``Factoring requirement dependencies in software requirement
  selection using graphs and integer programming,'' in \emph{2016 31st IEEE/ACM
  International Conference on Automated Software Engineering (ASE)}.\hskip 1em
  plus 0.5em minus 0.4em\relax IEEE, 2016, pp. 884--887.

\bibitem{mougouei2017dependency}
D.~Mougouei, D.~M. Powers, and A.~Moeini, ``Dependency-aware software release
  planning,'' in \emph{2017 IEEE/ACM 39th International Conference on Software
  Engineering Companion (ICSE-C)}.\hskip 1em plus 0.5em minus 0.4em\relax IEEE,
  2017, pp. 198--200.

\bibitem{mougouei2017modeling}
D.~Mougouei and D.~M. Powers, ``Modeling and selection of interdependent
  software requirements using fuzzy graphs,'' \emph{International Journal of
  Fuzzy Systems}, pp. 1--17, 2017.

\bibitem{shi2017new}
X.~Shi, L.~Wu, and X.~Meng, ``A new optimization model for the sustainable
  development: Quadratic knapsack problem with conflict graphs,''
  \emph{Sustainability}, vol.~9, no.~2, p. 236, 2017.

\bibitem{carlshamre_industrial_2001}
P.~Carlshamre, K.~Sandahl, M.~Lindvall, B.~Regnell, and J.~Natt~och Dag, ``An
  industrial survey of requirements interdependencies in software product
  release planning,'' in \emph{Fifth {IEEE} International Symposium on
  Requirements Engineering, 2001. Proceedings}, 2001, pp. 84--91.

\bibitem{dahlstedt2005requirements}
{\AA}.~G. Dahlstedt and A.~Persson, ``Requirements interdependencies: state of
  the art and future challenges,'' in \emph{Engineering and managing software
  requirements}.\hskip 1em plus 0.5em minus 0.4em\relax Springer, 2005, pp.
  95--116.

\bibitem{ngo_wicked_2008}
A.~Ngo-The and G.~Ruhe, ``\BIBforeignlanguage{English}{A systematic approach
  for solving the wicked problem of software release planning},''
  \emph{\BIBforeignlanguage{English}{Soft Computing}}, vol.~12, no.~1, pp.
  95--108, 2008.

\bibitem{ngo2005measuring}
A.~Ngo-The and M.~O. Saliu, ``Measuring dependency constraint satisfaction in
  software release planning using dissimilarity of fuzzy graphs,'' in
  \emph{Cognitive Informatics, 2005.(ICCI 2005). Fourth IEEE Conference
  on}.\hskip 1em plus 0.5em minus 0.4em\relax IEEE, 2005, pp. 301--307.

\bibitem{wang_simulation_2012}
J.~Wang, J.~Li, Q.~Wang, H.~Zhang, and H.~Wang, ``A simulation approach for
  impact analysis of requirement volatility considering dependency change,''
  \emph{Requirements Engineering: Foundation for Software Quality}, pp. 59--76,
  2012.

\bibitem{schwartz1987toward}
S.~H. Schwartz and W.~Bilsky, ``Toward a universal psychological structure of
  human values.'' \emph{Journal of personality and social psychology}, vol.~53,
  no.~3, p. 550, 1987.

\bibitem{zadeh_fuzzysets_1965}
L.~A. Zadeh, ``Fyzzy sets,'' \emph{Inf. Comput.}, vol.~8, pp. 338\--353, Dec
  1965.

\bibitem{mougouei2013fuzzy}
D.~Mougouei and W.~Nurhayati, ``A fuzzy-based technique for describing security
  requirements of intrusion tolerant systems,'' \emph{International Journal of
  Software Engineering and its Applications}, vol.~7, no.~2, pp. 99--112, 2013.

\bibitem{de1984qualitative}
J.~De~Kleer and J.~S. Brown, ``A qualitative physics based on confluences,''
  \emph{Artificial intelligence}, vol.~24, no.~1, pp. 7--83, 1984.

\bibitem{wellman1990formulation}
M.~P. Wellman and M.~Derthick, \emph{Formulation of tradeoffs in planning under
  uncertainty}.\hskip 1em plus 0.5em minus 0.4em\relax Pitman London, 1990.

\bibitem{kusiak_1995_dependency}
A.~Kusiak and J.~Wang, ``Dependency analysis in constraint negotiation,''
  \emph{Systems, Man and Cybernetics, IEEE Transactions on}, vol.~25, no.~9,
  pp. 1301--1313, 1995.

\bibitem{rosenfeld_fuzzygraph_1975}
A.~Rosenfeld, ``Fuzzy graphs,'' \emph{Fuzzy Sets and Their Applications},
  vol.~77, p.~95, 1975.

\end{thebibliography}
